\documentclass[a4paper,11pt]{article}

\usepackage{amssymb}

\begin{document}

\title{Security issues in a group key establishment protocol}
\author{Chris J Mitchell\\
  Information Security Group, Royal Holloway,
  University of London, UK
}

\date{16th March 2018 (v3)}
\maketitle

\begin{abstract}
Major shortcomings in a recently published group key establishment protocol are
described.  These shortcomings are sufficiently serious that the protocol
should not be used.
\end{abstract}

\section{Introduction}

Harn and Hsu \cite{Harn17} recently published a protocol designed to provide
authenticated group key establishment.  In this brief note we describe a number
of serious security issues with this scheme; in particular it does not provide
the properties claimed.

The remainder of the paper is structured as follows.  Section~\ref{protocol}
defines the protocol, including the intended context of use.
Section~\ref{analysis} then describes a number of serious issues with the
protocol.  The paper concludes in section~\ref{end}.

\section{The Harn-Hsu key establishment protocol} \label{protocol}

\subsection{Context and goals}  \label{goals}

The protocol is intended for use by a pre-established community of users, and
enables any subset (group) of this community to agree on a shared secret key,
where this secret key can be chosen and distributed by any member of the
community.  Group key establishment protocols have been widely discussed in the
literature for many years --- see, for example, chapter 6 of Boyd and Mathuria
\cite{Boyd03}. Indeed, the area is so well-established that an ISO/IEC standard
for group key establishment \cite{ISO11770-5:11} was published back in 2011.

The threat model for such protocols varies, but typically the goal is that,
after completion of the protocol, all participants agree on the same key, they
know it is `fresh', and that no parties other than those intended learn
anything about the key.

As far as the protocol described by Harn and Hsu \cite{Harn17} is concerned,
the following statements are made regarding its intended use and properties.
\begin{itemize}
\item `In our protocol, each member needs a pair of long-term DH
    [Diffie-Hellman] private and public keys and the long-term DH public
    key has been digitally signed by a trusted Certificate Authority (CA)'.
\item `The group key is determined by an initiator of the group
    communication and broadcasts the group key to all group members.  The
    initiator can be any member in a group communication.  Each group key
    is used for only one communication session.  When a new group
    communication session is established, a new group key will be generated
    by an initiator'.  From this statement (and the use of `long-term') it
    is clear that the DH private and public keys are intended for use to
    establish many group keys.
\item 'The digital certificate of public keys of group members will be used
    by an initiator to assure that the group key can only be decrypted by
    legitimate group members but not by any non-members'.  This establishes
    a key goal of the protocol, i.e.\ to ensure that the established key is
    only available to the parties intended by the initiator.
\end{itemize}

Section 2.2 of Harn and Hsu \cite{Harn17} (entitled `Types of attackers'),
describes the two classes of attacker against which the protocol is intended to
be robust, namely \emph{insider attackers} and \emph{outside attackers}.  The
paper states `The insider attacker is a legitimate member who knows the group
key \ldots [and] is able to impersonate other members in a secure group
communication'.  As we show in section~\ref{impersonation} below, precisely
such an insider attack is possible.  This contradicts the claim made
(\cite{Harn17}, section 2.2.2) that `none of these attacks can work properly
against our protocol'.

\subsection{Related work}  \label{related}

The Harn-Hsu protocol uses a combination of secret sharing and Diffie-Hellman
key agreement.  The use of secret sharing as part of a group key establishment
protocol is long-established (see, for example, section 6.7.2 of Boyd and
Mathuria \cite{Boyd03}).  However, this approach is known to have shortcomings;
in particular the following issue is described in \cite{Boyd03}.
\begin{quote}
However, when we look at the question of sending different session keys over
time there are some problems.  A malicious principal who obtains one key gains
information regarding the shares of other principals \ldots
\end{quote}
As we describe below, a related problem arises with the Harn-Hsu scheme.
Indeed, the fact that the Harn-Hsu protocol has serious flaws is hardly
surprising given the unfortunate history of the area.  Back in 2010, Harn and
Lin \cite{Harn10} described a group key transfer protocol based on secret
sharing which is not only mathematically flawed, but also possesses very
serious security issues; this gave rise not only to a number of papers pointing
out the flaws (see, for example, \cite{Nam11,Nam12}), but also to further
flawed protocols attempting to `fix' these flaws.  Some of the history of this
domain can be found in the recent paper of Liu et al \cite{Liu17}.

\subsection{The protocol}

The following requirements apply for use of the protocol.
\begin{itemize}
\item The protocol is designed to work within a set ${\cal
    U}=\{U_1,U_2,\ldots,U_n\}$ of $n$ users.
\item Integers $p$, $q$ and $g$ must be agreed by all members of ${\cal
    U}$, where $p$ is a large prime (1024 bits is suggested), $q$ is a
    prime factor of $p-1$ (160 bits is suggested), and $g$ ($1<g<p$) is a
    generator of $\mathbb{Z}_q$.  All participants must also agree on a
    one-way hash-function $h$.
\item Every user $U_i$ must:
     \begin{itemize}
     \item have a unique identifier $\mbox{ID}_i$ (an integer
         satisfying $0\leq\mbox{ID}_i\leq q-1$), and
     \item choose a Diffie-Hellman private key $x_i\in\mathbb{Z}_q$,
         and obtain a CA-signed certificate for the associated public
         key $y_i=g^{x_i}\bmod q$.
     \end{itemize}
\end{itemize}

Now suppose user $U_w$ wishes to act as an \emph{initiator}, and establish a
new secret key $K$ between the members of a group of users ${\cal U}'$ (${\cal
U}'\subseteq{\cal U}$).  Suppose ${\cal U}' =
\{U_{z_1},U_{z_2},\ldots,U_{z_\ell}\}$ for some $\ell$ ($1\leq\ell\leq n$),
where $1\leq z_i\leq n$ for every $i$ ($1\leq i\leq\ell$).

Observe that we have made two minor changes to the notation of \cite{Harn17} to
avoid possible confusion. Harn and Hsu refer to the initiator as $U_s$, but
they also use $s$ to denote a ephemeral secret known only to the initiator.
They refer to the members of the group ${\cal U}'$ as
$\{U_{r_1},U_{r_2},\ldots,U_{r_\ell}\}$, but they then use $r$ to denote a
function of the ephemeral secret $s$.

The initiator proceeds as follows.
\begin{enumerate}
\item The initiator selects a one-time (ephemeral) secret
    $s\in\mathbb{Z}_q$, and computes $r=g^s\bmod q$.
\item The initiator obtains trusted copies of the public keys $y_{z_i}$ of
    every member of ${\cal U}'$, e.g.\ by obtaining and verifying the
    relevant public key certificates, and for every $i$ ($1\leq i\leq
    \ell$) uses its own private key $x_w$ and the ephemeral secret $s$ to
    compute a one-time shared secret key
    \[ k_{z_i} = (y_{z_i}^{x_w+s}\bmod p)\bmod q. \]
\item The initiator uses Lagrange interpolation to determine a polynomial
    $f(x)$ of degree $\ell$ which passes through the following set of
    $\ell+1$ points:
    \[ \{ (0,K), (\mbox{ID}_{z_1},k_{z_1}), (\mbox{ID}_{z_2},k_{z_2}), \ldots,
    (\mbox{ID}_{z_\ell},k_{z_\ell}) \}\]
observing that the key $K$ is treated here as an integer in $\mathbb{Z}_q$,
i.e.\ the choice of $q$ constrains the length of the established key $K$.
\item The initiator chooses an arbitrary set $S=\{a_1,a_2,\ldots,a_\ell\}$
    of size $\ell$, where $a_i\in\mathbb{Z}_q$ for every $i$ and
    $S\cap{\cal U}'=\emptyset$, and computes the $\ell+1$ public values
    $(a_1,f(a_1)), (a_2,f(a_2)), \ldots, (a_\ell,f(a_\ell))$ and $h(t||K)$,
    where $t$ is a timestamp.
\item The initiator now broadcasts $r$, $t$ and the $\ell+1$ public values
    \[(a_1,f(a_1)), (a_2,f(a_2)), \ldots, (a_\ell,f(a_\ell)), h(t||K)\]
    to all members of ${\cal U}'$.
\end{enumerate}

On receipt of the broadcast, each user $U_{z_i}\in{\cal U}'$ ($1\leq
i\leq\ell$) proceeds as follows.
\begin{enumerate}
\item $U_{z_i}$ recomputes the one-time secret key (shared with the
    initiator) as:
    \[ k_{z_i}=((ry_w)^{x_{z_i}}\bmod p)\bmod q.\]
\item $U_{z_i}$ uses Lagrange interpolation to recompute the polynomial
    $f(x)$ of degree $\ell$, using the following set of $\ell+1$ points:
    \[ \{(\mbox{ID}_{z_i},k_{z_i}), (a_1,f(a_1)), (a_2,f(a_2)), \ldots, (a_\ell,f(a_\ell))\}.\]
    $U_{z_i}$ can now recover $K'=f(0)$.
\item $U_{z_i}$ verifies that the received timestamp $t$ is sufficiently
    recent, computes $h(t||K')$, and checks that this equals the received
    hash value.  If so, the recomputed key $K'$ is correct, i.e.\ $K'=K$,
    and can be used for group communication.
\end{enumerate}

\subsection{Security claims} \label{claims}

Amongst others, Harn and Hsu \cite{Harn17} make the following claims regarding
the security properties of the protocol.
\begin{enumerate}
\item The protocol provides key authentication.  The meaning of this is not
    made completely clear, but it would appear that (and following common
    use of the term) this means that the group member can verify that the
    key originates from the claimed initiator and that it is a `fresh' key,
    i.e.\ it was sent by the initiator at the time indicated in the
    timestamp $t$.
\item The security of the secret sharing encryption is unconditionally
    secure.
\end{enumerate}

\section{Analysis} \label{analysis}

We now describe a number of serious issues with the protocol, including cases
where the protocol does not satisfy the security properties claimed of it.

\subsection{Missing information}

We firstly observe that, apart from the abuses of notation observed above, the
specification is missing certain key elements, including the following.
\begin{itemize}
\item It is not explicitly stated that $\mbox{ID}_i$ must be an element of
    $\mathbb{Z}_q$.
\item The message broadcast by the initiator must contain both the
    identifier of the initiator and the identifiers of the members of the
    group ${\cal U}'$. If the latter was not the case, then every user in
    ${\cal U}$ would be obliged to attempt to obtain the key $K$, and will
    only discover they are not a member of the group ${\cal U}'$ when the
    hash comparison fails.  This would impose a very significant
    unnecessary computational load on the global user set.  Moreover, the
    intended recipients would not know which other users know the key,
    making its use problematic.
\end{itemize}

\subsection{Unconditional security}

It is claimed that `the security of the secret sharing encryption is
unconditionally secure' (see claim 2 of section~\ref{claims}).  However, it is
easy to see that the only part of the scheme which can be considered as in any
sense unconditionally secure is the reconstruction of $f$.  However, if the
discrete logarithm problem can be solved with respect to $g$ in $\mathbb{Z}_p$,
then clearly all user private keys can be obtained from their public keys,
meaning that anyone with access to the relevant public keys can obtain $K$ from
a broadcast. That is, in no sense is the encryption of $K$ unconditionally
secure.

\subsection{Effects of compromise of a group key}

Suppose a group key $K$ is compromised, i.e.\ it becomes available to a
malicious party $M$ (insider or outsider), who also has access to the
corresponding broadcast message, i.e.:
\[ r,t,h(t||K),(a_1,f(a_1)),(a_2,f(a_2)),\ldots,(a_\ell,f(a_\ell)).\]
$M$ can now, at any time, choose a current timestamp, $t'$ say, and compute $h(t'||K)$.
$M$ can now impersonate the initator and sent the slightly modified
broadcast message:
\[ r,t',h(t'||K),(a_1,f(a_1)),(a_2,f(a_2)),\ldots,(a_\ell,f(a_\ell)).\]

This will be accepted as valid by all the recipients of the original (valid)
broadcast, i.e.\ they will accept $K$ as a newly generated, authentic key. This
attack can be repeated as many times as $M$ wishes, i.e.\ $M$ can force
continued use of a compromised key indefinitely, breaking key authentication
(i.e.\ claim 1 of section~\ref{claims}).

\subsection{Impersonation of an initiator} \label{impersonation}

Suppose user $U_{z_i}$ is a valid recipient of a broadcast, i.e.\
$U_{z_i}\in{\cal U}'$; then, since $U_{z_i}$ can compute the polynomial $f(x)$
used in this broadcast, $U_{z_i}$ can also compute all the one-time secret keys
\[ k_{z_1}, k_{z_2}, \ldots, k_{z_\ell}\]
for members of the group ${\cal U}'$, simply by computing $f(z_j)$ for every
$j$ ($1\leq j\leq\ell$, $j\not=i$).

This information enables $U_{z_i}$ to impersonate the valid initiator in a
broadcast of a key chosen by $U_{z_i}$ to the original set of recipients (or
any subset of the original set of recipients) at any time.  The attack works in
the following way.
\begin{enumerate}
\item $U_{z_i}$ chooses a new key $K^*$ and a current timestamp $t^*$.
\item $U_{z_i}$ uses Lagrange interpolation to determine a polynomial
    $f^*(x)$ of degree $\ell$ which passes through the following set of
    $\ell+1$ points:
    \[ \{ (0,K^*), (\mbox{ID}_{z_1},k_{z_1}), (\mbox{ID}_{z_2},k_{z_2}), \ldots,
    (\mbox{ID}_{z_\ell},k_{z_\ell}) \}.\]
\item $U_{z_i}$ now chooses a set $S^*=\{a^*_1,a^*_2,\ldots,a^*_\ell\}$ of
    size $\ell$, where $a^*_i\in\mathbb{Z}_q$ for every $i$ and
    $S^*\cap{\cal U}'=\emptyset$, and computes the $\ell+1$ values
    $(a^*_1,f(a^*_1)), (a^*_2,f(a^*_2)), \ldots, (a^*_\ell,f(a^*_\ell))$
    and $h(t^*||K^*)$.
\item Finally $U_{z_i}$ impersonates the original initiator to broadcast
    $r$ (taken from the original valid broadcast), $t^*$ and the $\ell+1$
    values computed in the previous step to all members of ${\cal U}'$.
\item It is straightforward to verify that the broadcast will be accepted
    by all members of the group ${\cal U}'$.
\end{enumerate}

That is, at any time after the original broadcast, any of the recipients of the
broadcast can send a new broadcast message containing a new key and timestamp
to all the members of the original group, impersonating the original initiator.
This insider attack clearly breaks the key authentication property (i.e.\ claim
1 of section~\ref{claims}), and is also clearly something that the designers of
the protocol did not intend to be possible since, as discussed in
section~\ref{goals}, insider attackers are part of the Harn-Hsu threat model.

Note that this attack relates to the observation made by Boyd and Mathuria
\cite{Boyd03} regarding the security properties of key establishment protocols
based on secret sharing --- see section \ref{related}.

\section{Conclusions}  \label{end}

As demonstrated above, the protocol proposed by Harn and Hsu \cite{Harn17}
fails to possess the properties claimed of it.  This means that the protocol
should not be used.  It is important to observe that the Harn-Hsu paper does
not include a rigorous security proof using the state of the art `provable
security' techniques, nor is there a formal model of security for the protocol.
This helps to explain why fundamental flaws exist.  Indeed, the following
observation, made by Liu et al. \cite{Liu17} with respect to a number of
previously proposed but flawed group key establishment protocols, is hugely
pertinent.
\begin{quote}
The security proof for each vulnerable GKD protocol only relies on incomplete
or informal arguments.  It can be expected that they would suffer from attacks.
\end{quote}

It would, of course, be
tempting to try to repair the protocol to address the issues identified, but,
unless a version can be devised with an accompanying security proof, there is a
strong chance that subtle flaws will remain.  Certainly the analysis necessary
to find the flaws listed above was completed in a couple of hours, and no
attempt was made to discover all the possible attacks.


\end{document}